\documentclass[12pt,preprint]{aastex}
\usepackage[dvips]{epsfig,color}

\newcommand{\eg}{{\it e.g.,}}

\newcommand{\etal}{{\it et al.}}
\newcommand{\sersic}{S\'{e}rsic }
\newcommand{\drms}{\ensuremath{\Delta_{\rm rms}}}
\newcommand{\psd}{\ensuremath{\rho/\sigma^{3}}}

\newcommand{\ignore}[1]{\relax}

\shorttitle{Density Profiles}
\shortauthors{Barnes \etal}

\slugcomment{Rough draft, \today}

\begin{document}

\title{Density Profiles of Collisionless Equilibria. I. Spherical
Isotropic Systems}
\author{Eric I. Barnes, Liliya L. R. Williams}
\affil{Department of Astronomy, University of Minnesota, Minneapolis,
MN 55455}
\email{barnes,llrw@astro.umn.edu}
\author{Arif Babul\altaffilmark{1}}
\affil{Department of Physics \& Astronomy, University of Victoria, BC,
Canada}
\email{babul@uvic.ca}
\author{Julianne J. Dalcanton\altaffilmark{2}}
\affil{Department of Astronomy, University of Washington, Box 351580,
Seattle, WA 98195}
\email{jd@astro.washington.edu}

\altaffiltext{1}{Leverhulme Visiting Professor, Universities of Oxford
and Durham}
\altaffiltext{2}{Alfred P. Sloan Foundation Fellow}

\begin{abstract}

We investigate the connection between collisionless equilibria and the
phase-space relation between density $\rho$ and velocity dispersion
$\sigma$ found in simulations of dark matter halo formation, $F=\psd
\propto r^{-\alpha}$.  Understanding this relation will shed light on
the physics relevant to collisionless collapse and on the subsequent
structures formed.  We show that empirical density profiles that
provide good fits to N-body halos also happen to have nearly
scale-free \psd\ distributions when in equilibrium.  We have also done
a preliminary investigation of variables other than $r$ that may match
or supercede the correlation with $F$.  In the same vein, we show that
$\rho/\sigma^m$, where $m=3$ is the most appropriate combination to
use in discussions of the power-law relationship.  Since the
mechanical equilibrium condition that characterizes the final systems
does not by itself lead to power-law $F$ distributions, our findings
prompt us to posit that dynamical collapse processes (such as violent
relaxation) are responsible for the radial power-law nature of the
\psd\ distributions of virialized systems.

\end{abstract}


\section{Introduction}\label{intro}

The ``poor man's'' phase-space density proxy $F=\psd$, where $\rho$ is
density and $\sigma$ is total velocity dispersion, is a power-law in
radius ($F \propto r^{-\alpha}$) for a surprising variety of
self-gravitating, collisionless equilibria.  Isothermal systems have
$\rho \propto r^{-2}$ and constant $\sigma$, giving $\alpha=2$.  A
broader class of systems with power-law behavior in both $\rho$ and
$\sigma$ also naturally produce power-law behavior for $F$.  For
example, the self-similar collisionless infall models in
\citet[][\S4]{b85} have $\rho \propto r^{-9/4}$ and $\sigma \propto
r^{-1/8}$, leading to $\alpha=1.875$.  More surprising is that systems
in which neither $\rho$ nor $\sigma$ are power-laws can still possess
$F$ distributions that are.  For example, there is a growing body of
evidence, supported by results from simulations of increasingly higher
resolution and detail, that seems to suggest collisionless halos
formed in cosmological simulations are characterized by nearly
scale-free $F$ distributions, although they have decidedly
nonpower-law density profiles.  This was first noted by \citet{tn01}
who, at the time, determined that $\alpha=1.875$ over 3 orders of
magnitude in radius.  This value of $\alpha$, coincidently, is the
same as that derived by \citet{b85}.  More recent N-body simulations
have produced $\alpha$ values of; 1.95 \citep{r04}, 1.90 \citep{a04},
and 1.84 \citep[][based on the simulations in
\citet{d04a,d04b}]{dm05}.  \citet{a05} report that a very different,
semi-analytical halo formation method results in power-law $F$
distributions over similar radial ranges.  However, these authors find
a range of $\alpha$ values (including 1.875) that depend on initial
conditions.  As this formation method is much simpler than an N-body
evolution but still reproduces scale-free $F$, the physics responsible
for the distribution must be common to both techniques.  One such
process is violent relaxation.  In this work, we use ``violent
relaxation'' as shorthand for the incomplete relaxation process that
is due to the varying of potential during collapse rather than the
strict, complete relaxation discussed in \citet{l67}.  Also in the
\citet{a05} work, it is shown that in a totally isotropic system the
Jeans equation can be solved analytically and that there is a
``special'' $\alpha=35/18=1.94\bar{4}$.

It appears that power-law distributions of $F$ are robust features of
collisionless equilibria.  The exponents of the power-laws vary, but
cluster near values $\la2$.  This paper is part of a continuing series
of investigations aimed at understanding the ubiquity and the origin
of the phenomenon.  We specifically want to exploit its occurence to
gain insights into the processes governing the virialization of
collisionless halos.

In this paper, we review the conditions for hydrostatic equilibrium,
the Jeans equation.  By examining density profiles motivated by N-body
simulations and analyzing the associated $F$ distributions, in
\S\ref{power} we demonstrate that the Jeans equation by itself is not
sufficient to force a power-law for \psd.  At present, we restrict
ourselves to spherical equilibria with isotropic velocity
distributions.  Interestingly, typical density profiles that are used
to characterize data from cosmological N-body simulations all seem to
have nearly scale-free $F$ distributions, as do halos that are formed
semi-analytically \citep{a05}.  This aspect of N-body and
semi-analytically generated halos is certainly unexpected and
consequently, in \S\ref{constrain}, we investigate the implications of
explicitly imposing the requirement of scale-free \psd\ on the density
profiles of equilibrium structures.  We summarize our findings in the
final section.

\section{Empirical Density Profiles}\label{denprofs}

\begin{figure}
\plotone{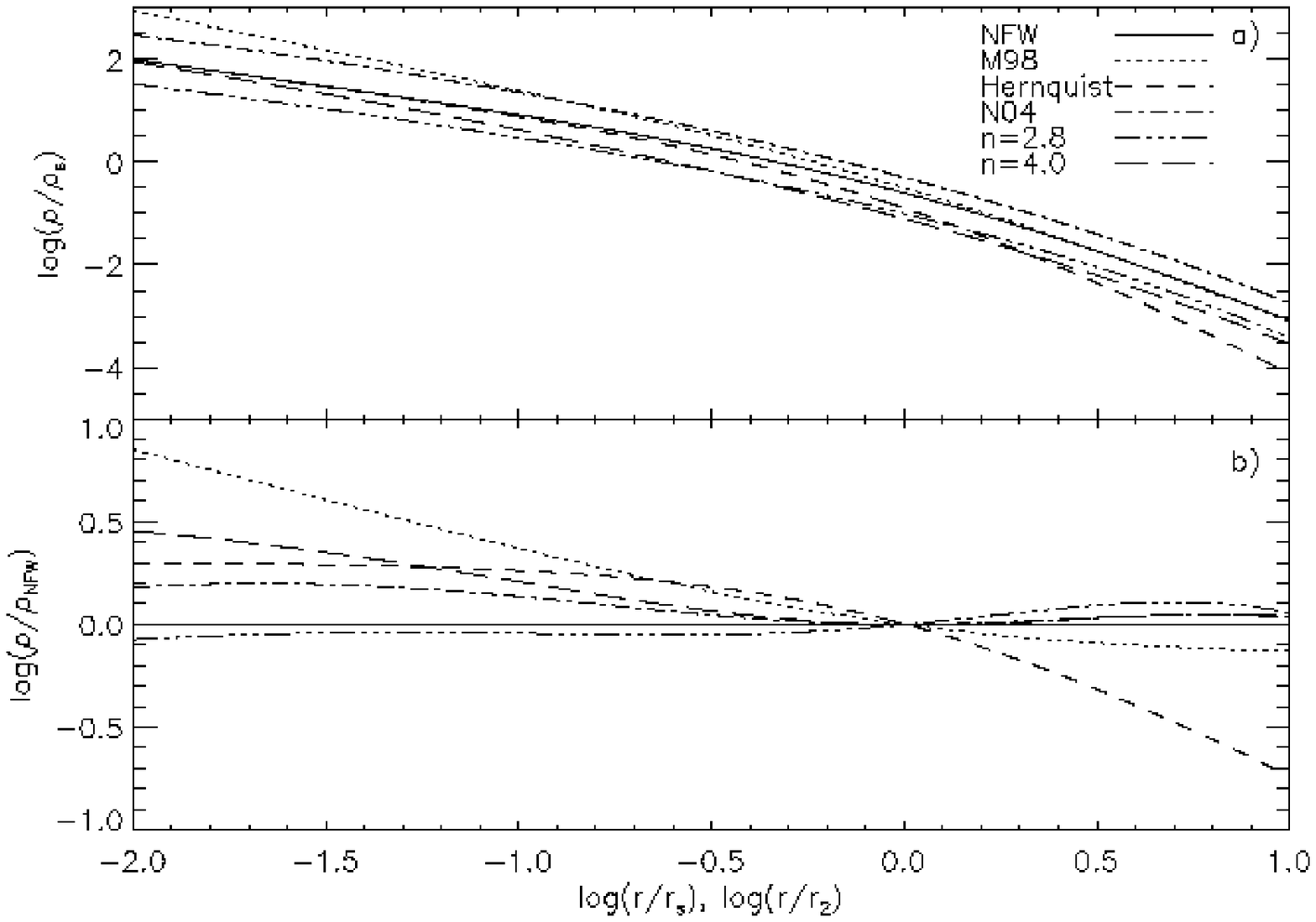}
\figcaption{Plots showing the NFW (solid line), \citet{m98} (dotted),
Hernquist (dashed), \citet{n04} (dash-dotted), \sersic $n=2.8$
(dash-triple dotted), and \sersic $n=4.0$ (long dashed) density
distributions.  The linestyles are the same in both plots.  a)  The
log-log density profiles.  The vertical normalization is arbitrary and
the curves have been somewhat separated to aid identification. b)  The
log-log density profiles of the same distributions divided the NFW
(horizontal solid line).  These ratio profiles have been normalized to
agree at the scale radius.
\label{denf}}
\end{figure}

In this work we will focus on several specific density profiles, shown
in Figures \ref{denf} \& \ref{halof}.  The standard
Navarro-Frenk-White \citep[][NFW]{nfw96,nfw97}, Moore et al
\citet[][M98]{m98}, and Hernquist \citep{h90} (solid, dotted, and
dashed lines, respectively) are examples of dual power-law
distributions with differing asymptotic behaviors that have been used
to fit density profiles of cosmological N-body halos.  The generalized
dual power-law profile has the form,
\begin{equation}\label{gnfw}
\frac{\rho}{\rho_s}=\left( \frac{r}{r_s}\right) ^{-c_1}
\left( 1+\frac{r}{r_s}\right) ^{-c_2},
\end{equation}
where $\rho_s$ and $r_s$ are a scale density and length, respectively.
The exponents $c_1$ and $c_2$ determine the asymptotic power-law
behavior of the profile; NFW: ($c_1=1,\ c_2=2$), M98: ($c_1=1.5=c_2$),
Hernquist: ($c_1=1,\ c_2=3$).  We define the negative logarithmic
density slope to be $\gamma \equiv
-d\log{(\rho/\rho_s)}/d\log{(r/r_s)}$.  For generalized dual
power-law profiles, the $\gamma$ distributions are given by,
\begin{equation}\label{nfwgam}
\gamma=\frac{c_1+(c_1+c_2)(r/r_s)}{1+(r/r_s)}.
\end{equation} 

The \citet{n04} profile (dash-dotted line) has been proposed as an
improvement over NFW for describing high resolution cosmological
N-body density profiles.  This profile never displays power-law
behavior; instead, the logarithmic density slope changes continuously
with $r$.  The expression that generates this curve is,
\begin{equation}\label{n04rho}
\ln{\left( \frac{\rho}{\rho_{2}}\right)}=
-\left( \frac{2}{\mu}\right) 
\left[ \left( \frac{r}{r_{2}}\right) ^{\mu}-1\right],
\end{equation}
where $r_{2}$ is the radius where $\gamma=2$ and $\rho_{2}$ is the
density at that radius.  The corresponding $\gamma$ profile is,
\begin{equation}\label{navgam}
\gamma=2 \left( \frac{r}{r_{2}} \right) ^{\mu}.
\end{equation}
As \citet{n04} found that $\mu=0.17$ best fit several N-body halo
profiles, we will refer to profiles ($\rho$ and $\gamma$) with
$\mu=0.17$ as N04 profiles, but we consider $0.001 \le \mu \le 0.22$.  

The final profile type we consider is the \sersic function
\citep{s68}.  The \sersic function is expressed analytically as,
\begin{equation}\label{sersicsig} \ln{\left(
\frac{\Sigma}{\Sigma_s}\right)}= -a_n \left[ \left( \frac{R}{R_s}
\right) ^{1/n}-1 \right],
\end{equation}
where $\Sigma$ is surface density, $R$ is projected distance, $n$
determines the shape of the profile, and $a_n$ is an $n$ dependent
constant chosen so that the projected mass interior to $R_s$ is equal
to the projected mass interior to $R=r_{2}$ for the N04 profile,
Equation~\ref{n04rho}.  This differs from the usual definition of the
\sersic constant that demands the projected mass within $R_s$ be half
the total mass.  Unfortunately, \sersic profiles do not readily
provide analytical expressions for $\rho$ or $\gamma$ [but see
\citet{t02} and \citet{g05}].  The dash-triple dotted and long dashed
lines in Figure~\ref{denf} show the calculated deprojected density
distributions for $n=2.9$ and $n=4.0$ (de Vaucouleurs profile),
respectively.  Larger (smaller) $n$ values reduce (increase) the
difference between the inner and outer logarithmic density slopes.
\citet{dh01} and \citet{m05} both suggest that the \sersic profile
describes the results of N-body simulations at least as well as the
previously discussed forms.  Further, \citet{dh01} point out that
$n\la4$ \sersic and NFW profiles have similar behaviors while
\citet{m05} find that $n\approx3$ provides the best fit to their dwarf
and galaxy-sized halos.

\section{\psd\ Distributions \& Equilibrium}\label{power}

Mechanical equilibrium for a spherical and isotropic collisionless
system is determined through the Jeans equation \citep{j19,bt87},
\begin{equation}\label{sjeans}
\frac{d}{dr}\left[\rho(r) \sigma^2(r)\right]=
-3G\rho(r)\frac{M(r)}{r^2},
\end{equation}
where $M(r)$ is the mass enclosed at radius $r$ and the factor of 3
comes from the definition
$\sigma^2=\sigma_r^2+\sigma_{\theta}^2+\sigma_{\phi}^2$ and the
isotropy of the system.  This equation certainly links $\rho$ and
$\sigma$, but does it alone impose power-law $F$ distributions?

\subsection{Specific Distributions}

\begin{figure}
\plotone{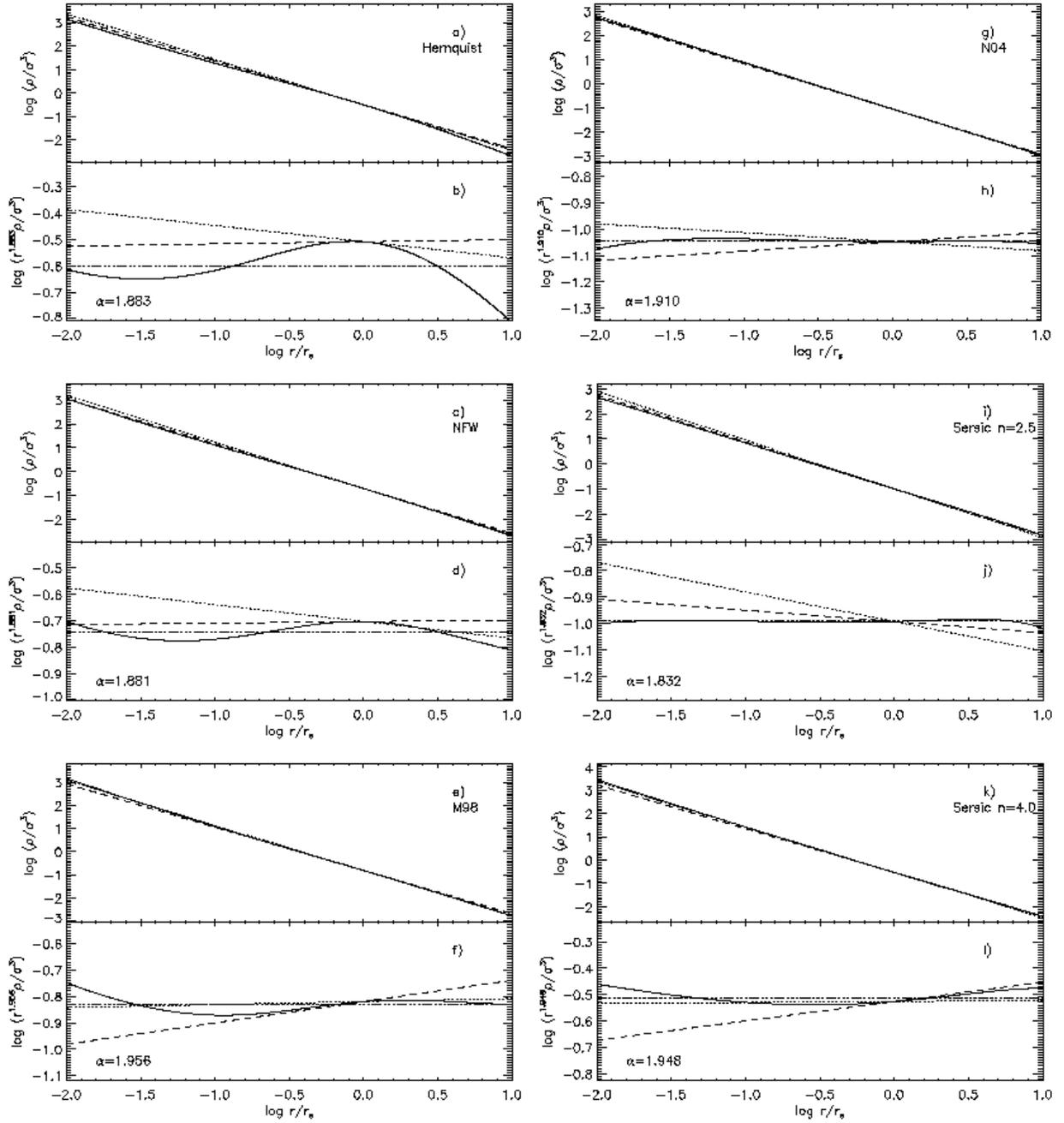}
\figcaption{Plots showing the raw $F$ distributions (panels a, c, e,
g, i, k) and versions scaled to highlight departures from a pure
power-law (panels b, d, f, h, j, l).  The dashed and dotted lines show
the behavior of power-law $F$ distributions with $\alpha=1.875$ and
$\alpha=35/18$, respectively.  The $\alpha$ values indicated in the
bottom panels are the slopes of the lines that best-fit the scaled
profiles.  Also in the bottom panels, the scaled best linear fits are
the dash-triple dotted horizontal lines, and the dotted and dashed
lines are the scaled power-laws corresponding to the lines in the
upper panels.  The density profiles are noted in the plots; Hernquist
(a,b), NFW (c,d), \citet{m98} (e,f), \citet{n04} (g,h), \sersic
$n=2.5$ (i,j), and \sersic $n=4.0$ (k,l).
\label{halof}}
\end{figure}

We demonstrate that the answer is no by providing a counter-example.
Inserting the Hernquist density profile into Equation~\ref{sjeans}, we
solve for $\sigma$ and thereby insure that the halo is in equilibrium.
The resulting $F$ distribution is shown as a solid line in
Figure~\ref{halof}, panels a and b.  In the top panels of this figure,
the dashed lines have slopes of -1.875, the dotted lines have slopes
of -35/18, and each line is normalized to the $F$ value at
$\log{(r/r_s)}=0$.  The curves in the bottom panels highlight
departures from the best-fit power-law behavior (horizontal
dash-triple dotted lines).  The dashed and dotted lines denote the
same power-laws as in the top panels, but scaled to the best-fit
slope.  We use $\alpha=1.875$ as a fiducial value because it is the
result of straightforward analytical calculation \citep{b85} as well
as being representative of the mean of the N-body results discussed in
the Introduction.  At the same time, we will also highlight the
analytically motivated $\alpha=35/18$ \citep{a05}.  The abscissa range
for the figure reflects that halos are usually resolved over roughly 3
orders of magnitude, from the virial radius [$\log{(r/r_s)}\approx 1$]
to about 1/1000 of the virial radius [$\log{(r/r_s)}\approx -2$].  The
dash-triple dotted lines in the bottom panels indicate the best linear
fits to the $F$ profiles.  The dotted and dashed lines represent the
same lines as in the top panel rescaled to the best linear fit slope.
The $\alpha$ values indicate the slope (modulo a minus sign) that the
best linear fit would have in the top panel.

We use the rms deviations \drms\ between the $F$ distributions and the
best power-law fits to quantify how close to a power-law each $F$ is.
These deviations are calculated over the resolved range of N-body halos,
from $\log{(r/r_s)}\approx -2$ to $\log{(r/r_s)}\approx 1$.  We adopt
the following convention for the rest of the paper; $F$ distributions
with $\drms \le 0.05$ will be considered power-laws, those with $\drms
>0.05$ will not.  This approximately reflects the level at which one
can detect a power-law by eye, \eg\ by looking at panel a.  With this
criterion, the Hernquist profile, with $\drms=0.07$, is evidence that
simple mechanical equilibrium does not enforce power-law $F$ behavior.
The Hernquist profile is not unique in this regard; King models
\citep[][not discussed in detail here]{k66} also produce $F$
distributions that have quite obvious deviations from power-law
shapes.

Having found these counter-examples, we now demonstrate that the other
empirical density profiles from \S\ref{denprofs} generally lead to
scale-free $F$ distributions.  In Figure~\ref{halof}, we present the
$F$ distributions calculated by solving Equation~\ref{sjeans} using
the NFW (panels c and d), M98 (e and f), and N04 (g and h) density
profiles.  These profiles have power-law $F$ distributions with $\drms
\la 0.03$ and $\alpha=1.881$, 1.956, and 1.910 for NFW, M98, and N04,
respectively.  N04 produces the best power-law $F$ distribution of
these 3 models, with $\drms =0.007$.  NFW and M98 profiles are poorer
(but still acceptable) power-laws with $\drms \approx 0.03$.  \sersic
profiles also produce power-law $F$ distributions, with the best
power-law $F$ ($n=2.5$, $\drms=0.005$, $\alpha=1.832$) shown in
Figure~\ref{halof} panels i and j.  We also include the results from
the de Vaucouleurs profile (\sersic $n=4.0$) in panels k and l.  This
range of $\alpha$ values (1.83-1.96) is approximately the same as the
range of results from N-body simulations (see the Introduction).
These findings are also in broad agreement with the results of
\citet{g05}.

\subsection{General Distributions}

\begin{figure}
\plotone{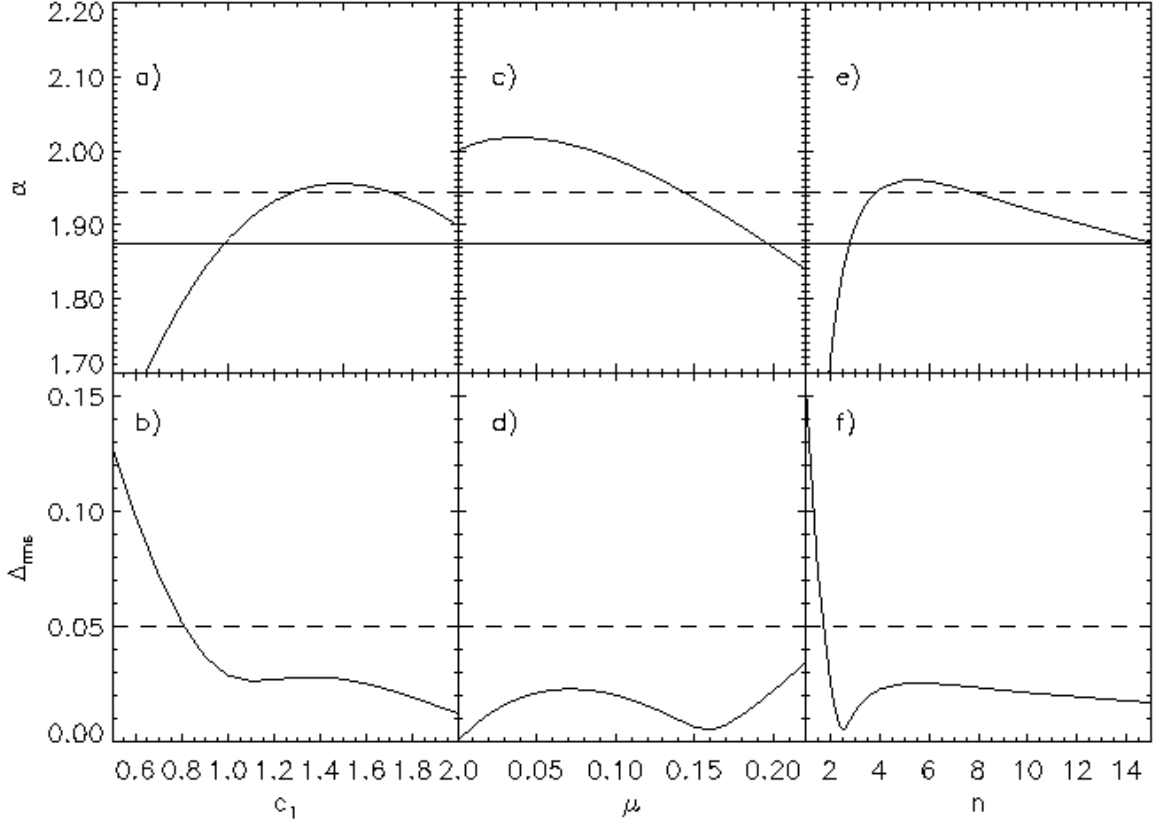}
\figcaption{Plot of the $\alpha$ and \drms\ values versus the shape
parameters for generalized dual power-law profiles (panels a \& b),
N04 profiles (panels c \& d), and \sersic profiles (panels e \& f).
The shape parameters are $c_1$ ($c_2=3-c_1$), $\mu$, and $n$ for the
dual-power law, N04, and \sersic profiles, respectively.  In the top
panels, the solid line lies at $\alpha=1.875$ and the dashed line
marks $\alpha=35/18$.  The dashed line in the bottom panels
illustrates the acceptable power-law cut-off value of $\drms=0.05$.
\label{pcomb}}
\end{figure}

\begin{figure}
\plotone{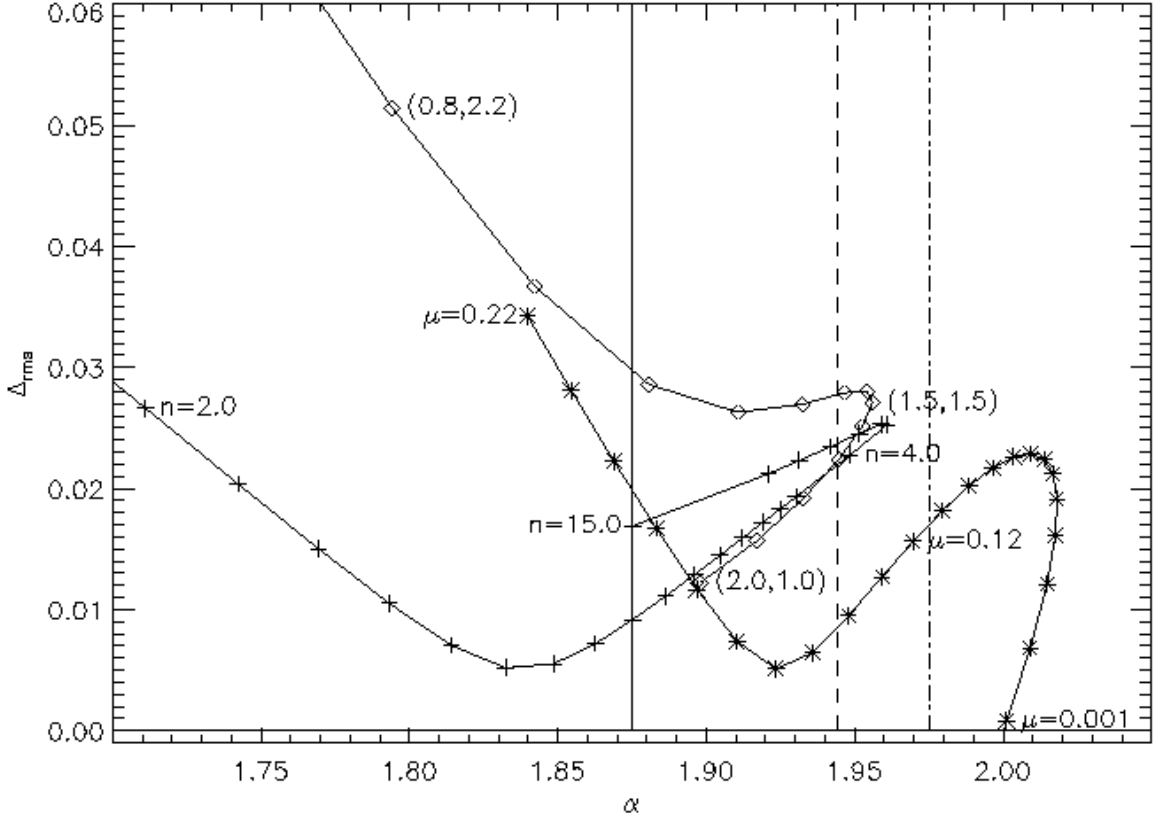}
\figcaption{$\alpha$ versus \drms\ for generalized
dual power-law (diamonds), N04 (asterisks), and \sersic (plus symbols)
profiles.  Along the \sersic track, the $n$ values increase from 2 to
15, with a turnaround point at $n=5$.  The $\mu$ values decrease from
left to right ($0.22 \rightarrow 0.001$) along the N04 track.  The
upper-leftmost diamond has ($c_1=0.8,c_2=2.2$), the diamond with the
largest $\alpha$ corresponds to the M98 profile ($c_1=c_2=1.5$), and
the diamond with the smallest \drms\ has ($c_1=2.0,c_2=1.0$).  The NFW
($c_1=1.0,c_2=2.0$) profile is marked by the diamond nearest to the
solid line at $\alpha=1.875$.  The dashed and dash-dotted lines mark
$\alpha=35/18$ and $\alpha=1.975$, respectively.
\label{arms}}
\end{figure}

In addition to these specific profiles, we have also examined the
generic forms of Equations~\ref{gnfw}, \ref{n04rho}, and
\ref{sersicsig}.  Varying the shape parameters ($c_1,c_2,\mu,n$) of
these profiles allows us to 1) find the profiles that have the best
power-law $F$ behavior and 2) determine the ranges of $\alpha$ values
that each profile supports.  We summarize the findings in
Figures~\ref{pcomb} and \ref{arms}.

Three classes of profiles are presented separately in
Figure~\ref{pcomb} to show the impact of the shape parameters on the
$F$ distributions.  Panels a and b display the results for generalized
dual power-law profiles with the constraint that $c_1+c_2=3$ (like NFW
and M98) and $0.5 \le c_1 \le 2.0$.  One can see that $\alpha=1.875$
is obtained when $c_1 \approx 1.0$ ($c_2 \approx 2.0$), very nearly
the canonical NFW profile.  However, the shallow local minimum in
panel b around $c_1=1.1$ indicates that the NFW profile does not give
the best power-law $F$ (for isotropic systems).\footnote{We point out
that all of these profile types can produce perfect power-law $F$
distributions ($\drms=0$) in the limit that the density becomes a
power-law; generalized dual power-law: $c_1 \rightarrow 3$, $c_2
\rightarrow 0$, \citet{n04}: $\mu \rightarrow 0$, and \sersic: $n
\rightarrow \infty$.  Since these pure power-law density profiles
result in unphysical infinite mass objects, we define the ``best''
power-law $F$ distributions to be determined by the local minima
apparent in the lower panels of Figure~\ref{pcomb}.}  The shallowness
of this minimum suggests that all values $1.0 \la c_1 \la 1.5$ give
similar quality power-law \psd\ distributions.  We note that the M98
profile ($c_1=1.5$) produces an $\alpha$ value closer to the
analytical value of 35/18, with the ($c_1=1.3$) case providing the
best fit to $\alpha=35/18$.  We have also investigated a few profiles
with $c_1+c_2=4$ and found that they do not form acceptable power-law
$F$ distributions.  Like the Hernquist profile ($c_1=1, c_2=3$), the
\drms\ values for these profiles are always $> 0.05$.  The \citet{n04}
profiles with $0.001 \le \mu \le 0.22$ give rise to panels c and d.
The $F$ distribution that produces the best power-law has $\mu \approx
0.16$, which is very close to the best-fit value $\mu=0.17$ from
\citet{n04}.  For $\mu \approx 0.14$, the corresponding $\alpha
\approx 35/18$.  This range in $\mu$ values is consistent with halos
found in the simulations of \citet{n04} $0.1 \la \mu \la 0.2$.  Among
\sersic profiles with $2.0 \le n \le 15.0$ (panels e and f), the model
at which \drms\ is minimum has $n=2.5$.  This $n$ value lies in the
range of values found in the \citet{m05} study.  Interestingly, the
\sersic profile that produces $\alpha=35/18$ has $n\la 4$, basically a
deVaucouleurs profile.

Pursuing this further, we turn to Figure~\ref{arms} which combines the
results from the three profile types by relating $\alpha$ and \drms\
values.  The plus symbols represent \sersic profile values, asterisks
mark \citet{n04} values, and diamonds show generalized dual power-law
values.  The vertical structure of this plot illustrates that the
\citet{n04} and \sersic profiles generally result in better power-law
$F$ distributions than the dual power-law form.  Interestingly, if we
think of the various simulation-inspired profiles in chronological
order (NFW, M98, and N04), it appears that the \psd\ distributions are
becoming better power-laws as the number of particles in simulations
increases, and the simulations themselves improve.  Such a trend may
be due to a decreased impact by two-body relaxation (which masks the
dynamics relevant to actual halos and decreases in importance with
increasing particle numbers) or it may be that the larger particle
numbers allow simulations to more faithfully reflect the pertinent
physics, \eg\ violent relaxation.

In the horizontal direction of Figure~\ref{arms}, we clearly see that
the profile types produce their best power-law at varying $\alpha$
values.  However, the minimum value of \drms\ for all the profiles
occur in a relatively narrow range of $\alpha$ values; between 1.84
and 1.97, close to the analytically derived value of
$\alpha=1.94\bar{4}$.  One thing to keep in mind is that this study
deals only with isotropic systems.  It could be that simulated N-body
halos, which have anisotropic velocity distributions \citep{hm04,b05},
are sufficiently different from these isotropic models to cause the
offsets.

\subsection{A More Fundamental Relation?}

\begin{figure}
\plotone{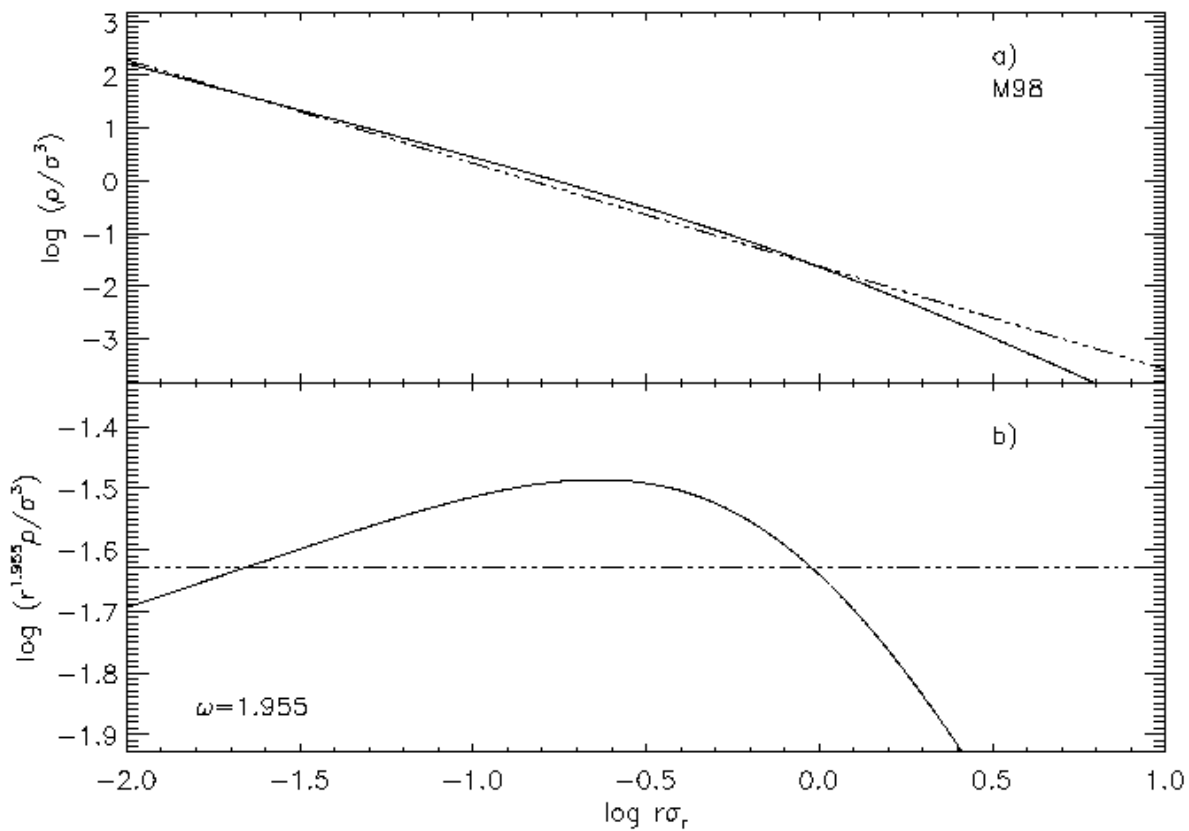}
\figcaption{Representative curves showing the relationship between
\psd\ and the radial action proxy $r \sigma_r$.  These specific curves
are for the M98 profile and should be compared to those in
Figure~\ref{halof} panels e and f.  Panel a shows the raw correlation
between $\log{F}$ and $\log{r \sigma_r}$ along with the best linear
fit (dash-triple dotted line).  Panel b displays the residuals between
the best-linear fit and the actual correlation.  The slope of the
best-fit line is given by the $\omega$ value.
\label{refp1}}
\end{figure}

The scale-free realtionship between $F$ and $r$ has been firmly
established, but we would like to know if there is a more dynamically
relevant quantity that shows a similar power-law correlation with $F$.
The list of candidate quantities is long, but we focus on two choices;
enclosed mass $M(r)$ and a proxy for the radial action $r \sigma_r$.
The $\log{F}$ vs. $\log{M}$ plots do not have power-law forms for any
of the distributions.  On the other hand, the $\log{F}$ vs.$\log{r
\sigma_r}$ curves do have approximately scale-free shapes, as shown in
Figure~\ref{refp1}a.  The best-fit line to this curve has a slope
($\omega=1.955$) that is very close the slope in Figure~\ref{halof}e
($\alpha=1.956$).  However, the comparison between the residuals shown
in Figure~\ref{refp1}b and those in Figure~\ref{halof}f demonstrate
that $F$ vs. $r$ is the better scale-free relation.  Indeed, the near
power-law relation between $F$ and $r \sigma_r$ occurs because
$\sigma_r$ has a very weak relation on $r$, making $F$ vs. $r
\sigma_r$ very similar to $F$ vs. $r$.

\citet{hmg04} find that a nontrivial function of potential accurately
describes the velocity dispersion profile in N-body halos.  Utilizing
a more general form of this function of potential, 
\begin{equation}
A=\Phi^a (\Phi_{\rm out}-\Phi)^b,
\end{equation}
we have investigated whether or not $F$ vs. $A$ provides a superior
power-law relation to $F$ vs. $r$.  We find that with appropriate
choices of $a$, $b$, and $\Phi_{\rm out}$, a power-law can be found
for $F$ vs. $A$ that is of comparable quality to that for $F$ vs. $r$.
However, we find that the degrees of freedom present in this function
allow it to closely resemble $r$ itself, making this function
unenlightening.  Despite the results of this brief search for a more
physically fundamental relation, we plan to continue investigating
alternative dynamical quantities.

\begin{figure}
\plotone{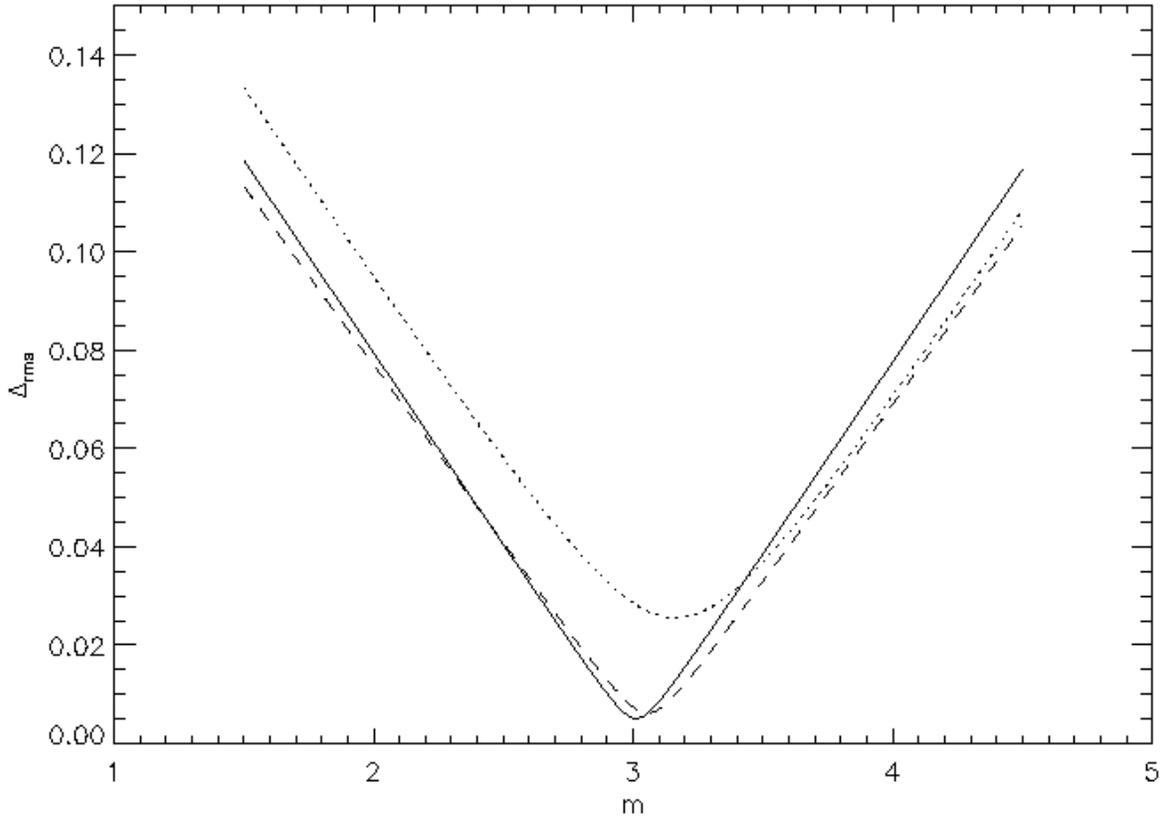}
\figcaption{Curves showing that the best $\rho/\sigma^m$ vs. $r$
power-law occurs when $m=3$.  The dotted line is the result of varying
$m$ for an NFW profile, while the dashed and solid lines illustrate
the variations for N04 and \sersic $n=2.5$ profiles, respectively.
\label{henfig}}
\end{figure}

One could also question whether or not our $F$ function is the most
illuminating choice of combination of $\rho$ and $\sigma$.  Certainly,
\psd\ is an interesting quantity, as it has the dimensions of
phase-space density, but would $\rho/\sigma^m$ work just as well
\citep{hpriv}?  For the NFW, N04, and \sersic functions, the answer
is no.  The deviations from a power-law distribution rapidly increase
as $m$ varies from 3 (over the interesting radial range $-2\le
\log{r/r_s} \le 1$).  This affinity for $m=3$ is obvious in
Figure~\ref{henfig} which shows the amplitude of the residuals from a
power-law $F$ vs. $r$ relationship as $m$ is varied from 1.5 to 4.5.

In this section we have shown that the condition of hydrostatic
equilibrium by itself does not produce power-law \psd.  However, the
density profiles that are used to fit the data from cosmological
N-body simulations all seem to have nearly scale-free \psd\
distributions, unlike the Hernquist and King profiles.  We have also
tried, in vain, to find more physically meaningful correlations
between $F$ and other quantities; enclosed mass, $r \sigma_r$, etc.
Furthermore, halos formed semi-analytically, through violent
relaxation \citep{l67}, also display $F \propto r^{-\alpha}$
\citep{a05}.  This aspect of N-body and semi-analytically generated
halos is certainly unexpected and prompts us to consider systems that
have explicitly scale-free \psd.

\section{The Constrained Jeans Equation}\label{constrain}

Imposing the constraint that $\psd=(\rho_0/v_0^3)(r/r_0)^{-\alpha}$
and using the dimensionless variables $x \equiv r/r_0$ and $y \equiv
\rho/\rho_0$, we rewrite Equation~\ref{sjeans} as,
\begin{equation}\label{sjeans2}
-\frac{x^2}{y}\left[ \frac{d}{dx}\left(y^{5/3}x^{2\alpha/3}
\right) \right]=BM(x),
\end{equation}
where $B=3G/r_0 v_0^2$.  Differentiating this equation with respect to
$x$ gives us,
\begin{equation}\label{sjeans3}
\frac{d}{dx}\left[-\frac{x^2}{y}\left\{
\frac{d}{dx}\left(y^{5/3}x^{2\alpha/3}
\right) \right\} \right]=Cyx^2,
\end{equation}
where $C=12\pi\rho_0 r_0^2/v_0^2$.  This expression is equivalent to
that presented in \citet{tn01}.  Following \citet{a05}, we eliminate
the constant $C$ by solving for $y$, differentiating with respect to
$x$ again, and grouping like terms.  The resulting constrained Jeans
equation is,
\begin{equation}\label{isojeans}
(2\alpha+\gamma-6)(\frac{2}{3}(\alpha-\gamma)+1)(2\alpha-5\gamma)=
15\gamma''+3\gamma'(8\alpha-5\gamma-5).
\end{equation}
In this notation, $\gamma=\gamma(x)=-d\ln{y}/d\ln{x}$ and the primes
indicate derivatives with respect to $\ln{x}$.

One way to connect power-law $F$ distributions and equilibria is by
making an analogy to fluid systems.  In hydrostatic equilibrium, the
term on the left-hand side of Equation~\ref{sjeans} is replaced by a
derivative of a single variable, the pressure $P$ (related to the
random motion in the system).  The important point is that $P$ is
related to $\rho$ through an equation of state.  This extra relation
closes the system of equations and, given boundary conditions, allows
one to solve for the equilibrium density distribution.  A power-law
$F$ distribution acts as a radius-dependent equation of state, linking
$\rho$ and the system's random motion, measured by $\sigma$.

\begin{figure}
\plotone{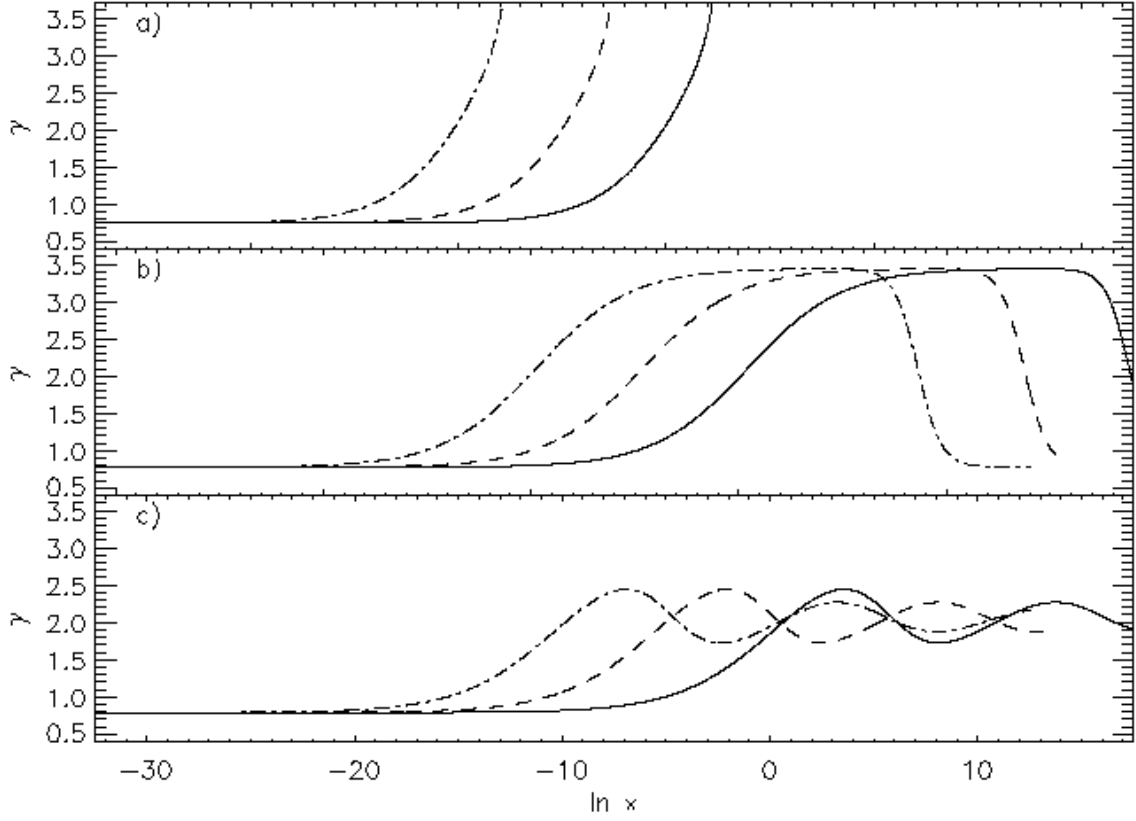}
\figcaption{Various $\gamma$ distributions that result from solving
Equation~\ref{isojeans}.  The top panel shows solutions with
$\alpha=1.875$.  Panel b utilizes $\alpha=35/18$, and the bottom panel
has $\alpha=1.975$.  The solid lines in each panel are solutions with
$\gamma'=5\times 10^{-6}$.  The dashed and dash-dotted lines have
$\gamma'=1\times 10^{-5}$ and $\gamma'=5\times 10^{-5}$, respectively.
The separations of the various profiles have been exaggerated for
clarity.
\label{jplot1}}
\end{figure}

\citet{a05} demonstrate that this equation has a rich set of solutions
that depend on the choices made for $\alpha$, initial $\gamma$
[$\gamma(0)$], and initial $\gamma'$ [$\gamma'(0)$].  In specific
density profiles, the asymptotic behavior of $\gamma(x)$ can be made
to increase without bound, to approach constant values, or even to
oscillate.  We show several types of solutions in Figure~\ref{jplot1}.
We choose $\gamma(0)$ through the relation $2\alpha-5\gamma(0)=0$,
representing a zero ``pressure'' derivative at the center \citep{a05}.
Once we choose an $\alpha$ value, the central $\gamma$ is set.  This
figure shows the impact of changing $\gamma'(0)$ from $5\times
10^{-6}$ to $5 \times 10^{-5}$ [see \citet{a05} for examples of
solutions with larger $\gamma'(0)$ values].  Smaller values of
$\gamma'(0)$ force the $\gamma(x)$ distribution to change less rapidly
with increasing $x$.  Since changing $\gamma'(0)$ values simply shifts
$\gamma$ distributions horizontally and does not affect the overall
shape (for $\gamma'(0) \la 10^{-3}$), we fix the $\gamma'(0)$ value
from here on.  Larger $\gamma'(0)$ values can make the $\gamma$
distribution convex in the region where $r=r_s$, unlike the
distributions that are of interest here.  Once $\gamma$ changes from
its initial value, the behavior is largely determined by $\alpha$.  As
mentioned earlier, $\gamma$ profiles display one of three kinds of
behavior and we choose to focus on three $\alpha$ values to provide
concrete examples; $\alpha=1.875$, $\alpha=35/18$ , and
$\alpha=1.975$.  $\alpha=35/18$ divides solutions in which $\gamma(x)$
increases indefinitely (like those in the top panel with
$\alpha=1.875$) from those in which $\gamma(x)$ acts as a damped
oscillator (bottom panel with $\alpha=1.975$).  A more extensive
discussion of this special $\alpha$ value can be found in
\citet[][\S3]{a05}.

In previous sections, we have discussed many types of density profiles
and have just shown that the constrained isotropic Jeans equation has
a wide variety of solutions.  We are now faced with the following
questions; which (if any) of the density profiles from \S\ref{power}
provides the best description of the constrained Jeans equation
solutions?, and does the answer to this question depend on the Jeans
equation parameter $\alpha$?

The thin solid lines in Figure~\ref{jplot2} are the $\gamma$ profiles
for the solutions of Equation~\ref{isojeans} with various $\alpha$
values and $\gamma'(0)=1\times 10^{-5}$.  The vertical solid lines
denote the radius at which $\gamma=2$ and the dotted vertical lines
mark $0.01$ and $10$ times this radius.  Panel a in
Figure~\ref{jplot2} has $\alpha=1.875$.  It is clear that the \sersic
form with $n=2.8$ provides a much better representation of the
solution curve than do the \citet{n04} profiles.  In panel b,
$\alpha=35/18$ and the quasi-asymptotic behavior of the solution curve
looks much more like one would expect for an NFW profile.  However, an
NFW $\gamma$ profile has a larger $\gamma'(r_s)$ than the solution
curve and does not provide a good approximation.  For this case,
neither the \sersic ($n=4.0$) nor the \citet{n04} curves are very good
matches to the solution.  The bottom panel shows the solution for
$\alpha=1.975$.  Again, the behavior of the solution curve is poorly
represented by either the \sersic ($n=5.0$) or \citet{n04} curves.
The \sersic behavior is not very surprising since Figure~\ref{arms}
shows that no \sersic profile can produce $\alpha$ values as large as
1.975.  While these last two plots point out the inadequacies of our
fitting functions for general solutions of the constrained Jeans
equation, in the case with $\alpha=1.875$, an average value from
N-body simulations, the \sersic profile provides a substanitally
better fit over the N04 profile to the solution of the isotropic
constrained Jeans equation.

\section{Summary \& Conclusions}

The apparent commonness of power-law distributions in the phase-space
density proxy $F\equiv \psd \propto r^{-\alpha}$ (density divided by
velocity dispersion cubed) in collapsed collisionless systems
\citep[\eg][]{tn01,a05} has lead us to investigate 1) whether or not
arbitrary equilibrium density profiles automatically lead to such
behavior and 2) the types of equilibria that occur under the
constraint that $F$ is scale-free.  In this study, we have only
investigated isotropic, spherically symmetric systems, but we will
soon extend this work to include anisotropic distributions.

We find the $F$ distribution corresponding to the Hernquist (or King)
profile is not an acceptable power-law and refutes the idea that
mechanical equilibrium alone is responsible for power-law $F$
distributions.  In general, profile types that empirically provide
good fits to N-body halos produce power-law $F$ behavior, with the
\citet{n04} and \sersic types being superior to the generalized dual
power-law profiles in this regard.  We speculate that this ubiquity is
not coincidence but rather that scale-free $F$ is a generic result of
the physics of collisionless collapse.  For the isotropic systems
considered here, the \sersic profile $F$ distributions with the
smallest \drms\ values tend to have smaller $\alpha$ values than those
corresponding to \citet{n04} profiles.  However, each type of
density profile covers the range of $\alpha$ values found in N-body
simulations.

\begin{figure}
\scalebox{0.5}{
\plotone{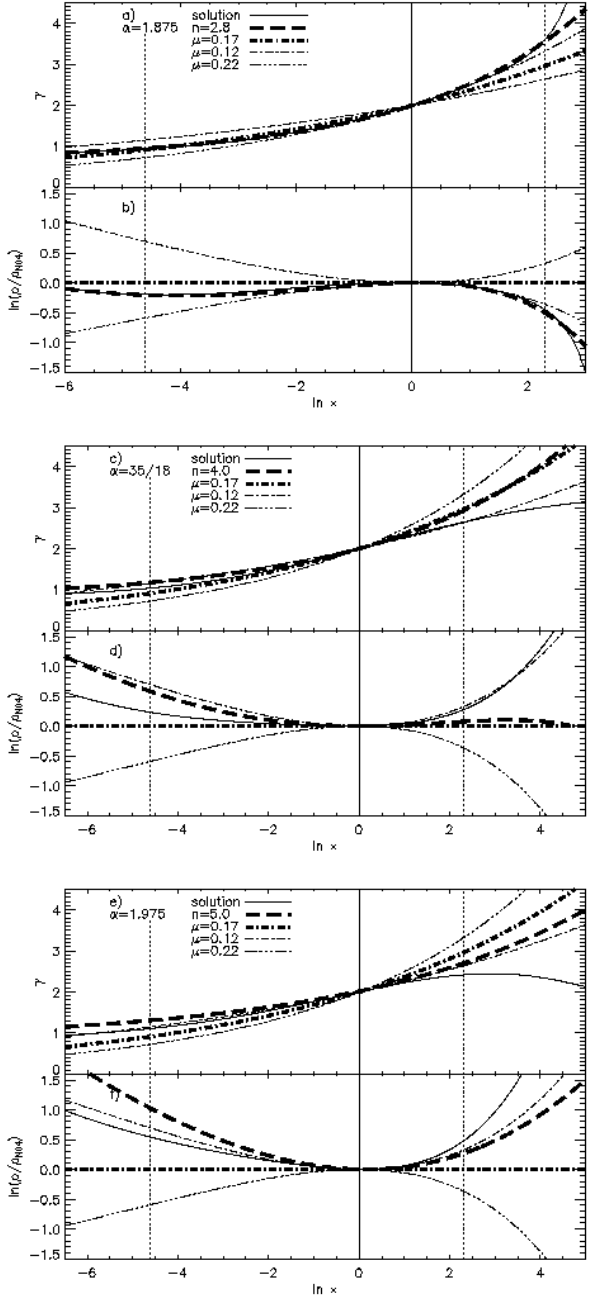}
}
\figcaption{Solutions of Equation~\ref{isojeans} with
$\gamma'(0)=1\times 10^{-5}$ (solid lines).  The top plots in each
panel show $\gamma$ distributions while the bottom panels present the
corresponding density profiles normalized by the N04 distribution.
The line types are the same in both plots.  The thick dashed and
dash-dotted lines correspond to \sersic and N04 density distributions,
respectively.  The thin dash-dotted lines are \citet{n04} profiles
with $\mu=0.12$; the thin dash-triple dotted lines have $\mu=0.22$.
Note that the range of $\ln{(x)}$ is much smaller than in
Figure~\ref{jplot1}; the vertical solid lines mark the positions of
$r_s$ or $r_{2}$; the dotted vertical lines are 1/100 and 10 times
this radius.  In plots a and b, $\alpha=1.875$ and the \sersic
$n=2.8$.  Plots b and c have ($\alpha=35/18, n=4.0$) and plots e and f
have ($\alpha=1.975, n=5.0$).
\label{jplot2}}
\end{figure}

Taking power-law $F$ behavior as a given allows us to write a
constrained Jeans equation that only involves the logarithmic density
slope $\gamma$, its derivatives, and $\alpha$.  This approach of
deriving equilibrium density (actually, $\gamma$) distributions and
comparing them to the $\gamma$ profiles corresponding to the
\citet{n04} and \sersic density profiles complements our earlier
findings.  The $\alpha=1.875$ results (top panel of
Figure~\ref{jplot2}) echo our previous conclusions that N-body halos
formed in cosmological simulations are best described by \sersic
models.

The preceeding points depend upon \psd\ vs. $r$ being the relevant
relationship.  As we do not have a compelling explanation for this
relation, we have also investigated other correlations of $F$ with
possibly more physical quantities.  These quantities have so far
failed to best the power-law correlation between $F$ and $r$.  Another
possibility is that $F$ itself is not the most illuminating variable.
Based on a thoughful suggestion from R. Henriksen, we have also looked
at whether or not the exponent of $\sigma$ in the combination
$\rho/\sigma^m$ can be changed to produce a better power-law.  Our
results clearly point to $m=3$ as the most interesting value.

One other question to ask is whether or not any type of relaxation to
equilibrium results in scale-free \psd.  In particular we have
wondered what effect two-body relaxation may have.  This is not to
suggest that current N-body simulations are affected by two-body
relaxation, but see \citet{ez05}.  One argument against the importance
of two-body relaxation in forming power-law $F$ is demonstrated by
King profiles, which accurately model two-body relaxed globular
clusters, but do not produce $F \propto r^{-\alpha}$.  This is
reminiscent of the findings of \citet{b82}.  That study found
significant differences between de Vaucouleurs and King models' $N(E)$
distributions ($N(E)dE$ is the number of particles with energies near
$E$).  Combining these findings with our own results for the King
profile as well as the results of \citet{a05} (halo formation without
any two-body effects produces scale-free $F$) brings us to conclude
that relaxation effects other than two-body interactions are
responsible for the power-law $F$ distributions.

We have demonstrated that, in equilibrium, density profiles that
accurately describe the end results of simulated collisionless
collapses (and hence violent relaxation) produce power-law $F$
distributions, while those that have been designed mostly for
analytical tractibility (\eg\ Hernquist profiles) or to describe
systems significantly different than galaxies (\eg\ King models) do
not.  And though there is no general theory explaining power-law $F$
behavior, our findings encourage us to speculate that dynamical
collapse processes (violent relaxation in particular) are playing a
major role in making \psd\ of equilibrium systems scale-free.

\acknowledgments
This work has been supported by NSF grant AST-0307604.  Research
support for AB comes from the Natural Sciences and Engineering
Research Council (Canada) through the Discovery grant program.  AB
would also like to acknowledge support from the Leverhulme Trust (UK)
in the form of the Leverhulme Visiting Professorship at the
Universities of Oxford and Durham.  JJD was partially supported
through the Alfred P. Sloan Foundation.  We would like to thank
Alister Graham and an anonymous referee for several helpful comments.
Extra thanks to Dick Henriksen for motivating us to investigate
variations in $F$.

\end{document}